\documentclass[aps,pre,twocolumn,groupedaddress,floatfix]{revtex4-1}
\usepackage{amsmath}
\usepackage{amssymb}
\usepackage{amsfonts}
\usepackage{graphicx}
\usepackage{bbold}
\usepackage{bm}

\newcommand{\avER}[1]{\left \langle #1 \right\rangle_{\text{ER}}}
\newcommand{\avW}[1]{ \left\langle #1 \right\rangle_{y}}

\newcommand{\bC     }{\mbox{\boldmath$C$}}

\DeclareMathOperator\arctanh{arctanh}
\DeclareMathOperator\sech{sech}

\usepackage{bm}
\usepackage{times,float}
\usepackage{graphicx}
\usepackage[usenames,dvipsnames,svgnames]{xcolor}
\usepackage{hyperref}
\hypersetup{colorlinks=true, linkcolor=NavyBlue, citecolor=PineGreen,urlcolor=cyan}

\begin{document}
\title{Phase transitions in atypical systems induced by a condensation transition on graphs}
\author{Edgar Guzm\'an-Gonz\'alez}
\email[]{edgar.guzman@fisica.unam.mx}
\affiliation{Department of Quantum Physics and Photonics, Institute of Physics, UNAM, P.O. Box 20-364, 01000 M\'exico City, M\'exico} 
\affiliation{London Mathematical Laboratory, 18 Margravine Gardens, London W6 8RH, United Kingdom}

\author{Isaac P\'erez Castillo}
\email[]{isaacpc@fisica.unam.mx}
\affiliation{Department of Quantum Physics and Photonics, Institute of Physics, UNAM, P.O. Box 20-364, 01000 M\'exico City, M\'exico} 
\affiliation{London Mathematical Laboratory, 18 Margravine Gardens, London W6 8RH, United Kingdom}

\author{Fernando L. Metz}
\email[]{fmetzfmetz@gmail.com}
\affiliation{Physics Institute, Federal University of Rio Grande do Sul, 91501-970 Porto Alegre, Brazil}
\affiliation{London Mathematical Laboratory, 18 Margravine Gardens, London W6 8RH, United Kingdom}

\date{\today}

\begin{abstract}
Random graphs undergo structural phase transitions that are crucial for dynamical processes and cooperative behavior of models defined on graphs. In this work we investigate the impact of a first-order structural transition on  the thermodynamics of the Ising model defined on  Erd\H{o}s-R\'enyi random graphs, as well as on the eigenvalue  distribution of the adjacency matrix of the same graphical model. The structural transition in question yields graph samples exhibiting condensation, characterized by  a large number of nodes having degrees in a narrow interval.  We show that this condensation transition induces distinct thermodynamic first-order transitions between the paramagnetic and the ferromagnetic phases of the Ising model. The condensation transition also leads to an abrupt change in the global eigenvalue statistics of the adjacency matrix, which renders the second moment of the eigenvalue distribution discontinuous. As a side result, we derive the critical line determining the  percolation transition in Erd\H{o}s-R\'enyi graph samples that feature condensation of degrees.
\end{abstract}

\maketitle

%%%%%%%%%%%%%%%%%%%%
\section{Introduction}
\label{sec:introduction}
%%%%%%%%%%%%%%%%%%%%
Random graphs are formidable tools to tackle problems in various disciplines, including physics, biology, and information science \cite{Newman10,Mezard09}. Informally speaking, a random graph is a collection of points or nodes interconnected by edges following a random prescription. In one of the simplest random graph models, each pair of nodes is connected (or not) according to a fixed probability, independently of the other nodes in the graph. This model is referred here as the Erd\H{o}s-R\'enyi (ER) random graph model, since it became popular after the seminal works by Paul Erd\H{o}s and Alfr\'ed R\'enyi \cite{Erdos59,Erdos60}.

There are essentially two main reasons why ER random graphs are useful mathematical models. On the one hand, due to their finite coordination number, ER random graphs arise naturally in problems that are described in terms of sparse interacting elements, where one unit is coupled to a finite number of others, such as in network theory \cite{Newman10} or in the solution of optimization problems \cite{Mezard09}.  On the other hand, ER random graphs can be seen as the infinite dimensional limit of Euclidean lattices. This property has led to analytic progress in the study of certain phase transitions that are otherwise very difficult to tackle in a finite-dimensional Euclidean space, such as the spin-glass transition \cite{Viana1985,Kanter1987,Monasson1998} and the Anderson localization transition \cite{Fyodorov91,Mirlin91,Slanina2012,Mata2017}. The absence of short loops and of any notion of Euclidean distance are distinctive features that allow for a mean-field description of systems interacting through the edges of ER random graphs.

Interestingly, random graphs undergo structural phase transitions when certain global statistical properties characterizing the graph structure change as a function of the model parameters \cite{Doro08,Palla04,Callaway2000,Cohen2001,Albert2001,Bianconi2018,Krap00,Bianconi01,Doro05,Park04a,Park04b,Annibale15,Strauss86,Burda04a,Burda04b,Gorsky2016,Gorsky2019,Coolen2019}. The percolation and the condensation transitions are emblematic examples in this context \cite{Doro08,Palla04}. In the first case, the largest connected component of a graph increases as a function of the mean number $c$ of neighbors per node. The graph percolates at a  critical value $c = c_{*}$, which means that, for $c \geq c_{*}$, the largest connected component contains a finite fraction of the total number of nodes. Percolation is a powerful notion to analyze the resilience of networks to random or targeted attacks \cite{Callaway2000,Cohen2001,Albert2001,Bianconi2018}, since the survival of the giant component with respect to the removal of a fraction of nodes is taken as an indication of network robustness. Random graphs undergo a condensation transition when a large number of subgraphs clump together to form a densely connected cluster. Different types of condensed graph configurations are possible, depending on the elementary structures composing the cluster. The simplest type of aggregate is formed through a phenomenon referred to as  condensation of edges \cite{Krap00,Bianconi01,Doro05}, when a finite fraction of the total number of edges attaches to a single node. Other examples include condensation of two-stars \cite{Park04a,Park04b,Annibale15} and triangles \cite{Strauss86,Burda04a,Burda04b}, where the elementary structures forming the cluster are paths of length two and cycles of length three, respectively.

More recently, reference \cite{Isaac_Fernando} has reported a novel type of structural phase transition in ER random graphs, characterized by an abrupt change in the degree statistics of the graph. The degree $K_i$ of a node $i$ is a random variable that counts the number of nodes connected to $i$. By varying a control parameter that allows to probe rare regions of the graph ensemble space, the degree distribution changes discontinuously from a Poisson form, typical of ER random graphs, to a distribution exhibiting a pronounced peak. This peaked distribution identifies a novel type of condensed state, where the degrees assume values in a narrow domain of its available configuration space. The formation of such condensed configurations has been coined {\it condensation of degrees}. These are large deviation events triggered by atypical fluctuations in the graph structure, similar to other random systems that exhibit condensation transitions driven by rare fluctuations \cite{zanetti2014,Corberi19}.

The influence of the graph structure on dynamical processes and on the cooperative behavior of models defined on random graphs is a key topic in network theory, which has been attracting a huge interest in recent decades \cite{Pastor2001,Pastorbook,Leone2002,Doro2002,Noh2004,Ichi2004,Perez2004, Perez2005,Perez2005b,Perez2005c,Lee2005,Doro08,Neri2016,Neri2019,lupo2019}. The degree statistics plays a pivotal role on the long-time behavior of random walks on graphs \cite{Noh2004}, on the critical threshold for epidemic spreading \cite{Pastor2001,Pastorbook}, on the linear stability of large interacting systems \cite{Neri2016,Neri2019}, and on the critical properties of cooperative systems defined on random graphs, such as the Ising model \cite{Leone2002,Doro2002}, the Kuramoto model \cite{Ichi2004,Lee2005,Perez2005b,Perez2005c,lupo2019}, and the classical Heisenberg model \cite{Perez2005c,lupo2019}. Since condensation of degrees emerges through a discontinuous transition in the degree distribution,  it is therefore compelling to ask how this structural transition impacts  the macroscopic behavior of systems interacting through the edges of random graphs.

Building on previous works on the large deviation theory of observables defined on graphs \cite{Metz_PerezC2017,PerezC_Metz2018,PerezC_Metz2018_2}, here we investigate how condensation of degrees influences two different problems: the thermodynamic phase transitions of the Ising model on an ER random graph and the eigenvalue distribution of the adjacency matrix of the graph. In the first case, large deviations in the graph structure, leading to condensation of degrees, induce different thermodynamic phase transitions, which are otherwise absent if one is limited to small, typical graph fluctuations. In fact, by computing the magnetization, the internal energy, and the magnetic susceptibility, we show that the Ising model displays three additional first-order transitions: a transition between ferromagnetic phases, a transition between paramagnetic phases, and a transition between a ferromagnetic and a paramagnetic phase. All these transitions are caused by the discontinuous change of the degree statistics. In our second example, we show that the eigenvalue statistics of the adjacency matrix of ER random graphs exhibits a discontinuous behavior across the condensation transition. In particular, the second moment of the eigenvalue distribution drops abruptly, indicating a concentration of eigenvalues around zero. These results are in contrast, for instance, with the percolation transition, where the eigenvalue distribution is insensitive to the formation of a giant component \cite{Bauer2001}. Incidentally, we also derive the percolation transition of atypical configurations of ER random graphs characterized by condensation of degrees, complementing the phase diagram presented in \cite{Isaac_Fernando}.

The paper is organized as follows. In the next section we define the ER random graph model and we introduce the main quantities to characterize the condensation transition in the graph structure. Section \ref{IsingXX} presents the results for the thermodynamics of the Ising model defined on rare samples of ER random graphs. The results for the eigenvalue distribution of the adjacency matrix of atypical configurations of ER random graphs are discussed in section \ref{spectrumXX}. We summarize our work and discuss some open problems in section \ref{conclusionXX}. Finally, two appendices provide detailed explanations of the analytical calculations for the Ising model and for the eigenvalue distribution.

%%%%%%%%%%%%%%%%%%%%
\section{Condensation of degrees}
%%%%%%%%%%%%%%%%%%%%
Erd\H{o}s-R\'enyi random graphs \cite{erdos} are simple undirected graphs with $N$ nodes, where the probability that two nodes are connected is $c/N$, with $c=O(1)$ independent of $N$. A single graph instance is completely defined through its $N \times N$ adjacency matrix $\bC$. The entry $c_{ij}$ of $\bC$ is one if node $i$ is connected to node $j$, and zero otherwise. The elements of $\bC$ are independent and identical distributed random variables drawn from the joint distribution
\begin{equation}
	P_{\text{ER}}(\boldsymbol{C})
        =\prod_{i < j} \left[ \frac{c}{N} \delta_{c_{ij},1}+\left(1-\frac{c}{N}\right)\delta_{c_{ij},0} \right] \,.
        \label{klla}
\end{equation}
The degree $K_i$ of a node $i$, defined as $K_i = \sum_{j =1 (\neq i)}^N c_{ij}$, is a random variable that counts the number of nodes connected to $i$. In the limit $N\rightarrow \infty$, the distribution of degrees $K_1,\dots,K_N$ becomes Poissonian with mean $c$
\begin{equation}
  p_c(k)=\frac{e^{-c} c^k}{k!}\,.
  \label{pois}
\end{equation}

In order to understand the meaning of condensation of degrees, it is useful to picture the nodes as particles and the different possible values of the degrees as energy levels. Thus, it is natural to ask how the total number of particles is distributed among the different energy levels. Condensation of degrees occurs when a large fraction of nodes (particles) is distributed over a few degrees (energy levels). Such phenomenon is captured by considering the random variable  $F_{[a,b]}(\bC)$ that counts the fraction of nodes having degrees in a certain interval $[a,b]$
\begin{equation}
F_{[a,b]}(\bC)=\frac{1}{N}\sum_{j=1}^N I_{[a,b]}(K_j)\,,
\end{equation}
where $I_{[a,b]}(x)$ is an indicator function, that is, $I_{[a,b]}(x)=1$ if $x\in [a,b]$, and zero otherwise. By computing the cumulant generating function of $F_{[a,b]}(\bC)$
\begin{equation}
\mathcal{G}(y) = \lim_{N \rightarrow \infty} \ln{\left\langle e^{y N F_{[a,b]}\left( \bC \right)}   \right\rangle_{\rm ER}}\,,
\end{equation}  
with $\avER{(\cdots)}$ denoting the average with the distribution $P_{\text{ER}}(\bC)$, reference \cite{Isaac_Fernando} has shown that the degree distribution changes abruptly from its typical Poissonian behavior, given by Eq. \eqref{pois}, to a peaked distribution. The latter distribution characterizes the formation of a condensed state, since a large fraction of nodes has similar degrees. The condensation transition is marked by a discontinuity of the first derivative $\frac{d \mathcal{G}(y)}{d y}$, which is the signature of a first-order phase transition in the parameter space $(c,y)$. The formation of the condensed state is a rare statistical event, triggered by large deviations in the graph structure, which produces two non-analytic points in the rate function controlling the large deviation probability \cite{Isaac_Fernando}.

There is an alternative way to interpret the problem that sheds light on the role of the parameter $y$. Instead of looking at the condensation transition from the viewpoint of large deviation theory, one can introduce a modified or constrained ER ensemble, in which the standard distribution $P_{\text{ER}}(\bC)$ is deformed by a Boltzmann-like weight that couples the external control parameter $y$ to the random variable $F_{[a,b]}(\bC)$. In this setting, the probability of drawing a graph with adjacency 
matrix $\bC$ is
\begin{equation}
	P_y(\bC)= \frac{P_{\text{ER}}(\bC) e^{y N  F_{[a,b]} \left( \bC \right) }}{\avER{e^{y N  F_{[a,b]} \left( \bC \right) }}}\,.
\label{eq:probWeightedY} 
\end{equation}
The role of $y$ becomes clear from Eq. \eqref{eq:probWeightedY}. If $y=0$, the weighted distribution $P_y(\bC)$ coincides with $P_{\text{ER}}(\bC)$. For positive (negative) values of $y$, the Boltzmann-like weight favors graphs where $F_{[a,b]}(\bC)$ is larger (smaller) than its typical value. Thus, $y$ is an external control parameter that biases the graph configurations and enables to probe the ensemble space of ER random graphs away from the typical configurations generated by Eq. \eqref{klla}. We can also interpret a change in $y$ as resulting from an external protocol to modify the graph structure: an increase (decrease) of $y$ corresponds to a rewiring of the links such that more (less) nodes have degrees in $[a,b]$.

The average of  any observable $A(\bC)$ over atypical ER graph configurations, conditioned by the value of $y$ through Eq. \eqref{eq:probWeightedY}, is thus obtained from
\begin{equation}
  \langle A(\bC) \rangle_y =  \lim_{N \rightarrow \infty} \frac{\avER{A(\bC)  e^{y N  F_{[a,b]}\left( \bC \right) }} }{\avER{ e^{y N  F_{[a,b]}\left( \bC \right)}} }\,.
  \label{klll}
\end{equation}
Our aim here is to study the impact of the first-order condensation transition on two paradigmatic problems defined on ER random graphs: the magnetic properties of the Ising model and the eigenvalue distribution of the adjacency matrix. Since in both examples we need to evaluate the ensemble average of certain observables that depend on $\bC$, Eq. \eqref{klll} provides a suitable starting point to obtain the typical properties of these systems constrained to rare sectors of the ER graph configuration space.

%%%%%%%%%%%%%%%%%%%%
\section{Ising model on constrained random graphs} 
\label{IsingXX}
\subsection{Model definitions and the free energy}
\label{sub:ising_model_in_a_random_graph}
%%%%%%%%%%%%%%%%%%%%
The Ising model is a mathematical model  to study the magnetic properties of a system. Here we are interested in the behavior of the Ising model on an ER random graph, i.e., the spin variables interact ferromagnetically through the edges of the graph. In particular, we will discuss the effect of the condensation transition, summarized in the previous section, on the magnetic properties of the Ising model.

Given a graph generated from the weighted ensemble defined by Eq. \eqref{eq:probWeightedY},  the energy of a configuration of binary spins $\bm{\sigma}=(\sigma_1,\dots,\sigma_N)$, with $\sigma_i \in \{-1,1\}$, is given by
\begin{equation}
	H_{\tiny{\bC}}(\bm{\sigma})=-J \sum_{i<j} c_{ij} \sigma_i \sigma_j
	-h \sum_i \sigma_i \, ,
\end{equation}
where $J > 0$ is the ferromagnetic coupling between any pair of adjacent spins, and $h$ is an external magnetic field. In the canonical ensemble, the thermodynamical properties are captured by the intensive free energy
\begin{equation}
	f(\bC)=- \frac{1}{\beta N} \ln Z(\bC)\,,
\end{equation}
where $\beta= 1/T$ is the inverse temperature of the system (the Boltzmann constant is equal to one), and $Z(\bC)$ is the partition function of the Ising model for a single realization of the graph
\begin{equation}
	Z(\bC)=\sum_{\bm \sigma}e^{-\beta H_{\tiny{\bC}}(\bm \sigma)} \,.
	\label{eq:partitionFunctionDef} 
\end{equation}
By assuming that, for a fixed value of $y$, the intensive free energy is a self-averaging quantity in the limit $N \rightarrow \infty$, the thermodynamics of the model is determined by the ensemble average of $f(\bC)$ over the graph configurations in the constrained ensemble
\begin{equation}
	f= - \lim_{N \rightarrow \infty} \frac{1}{\beta N}\frac{\avER{\ln Z(\bC) e^{y N  F_{[a,b]} \left( \bC \right) }}}{\avER{e^{y N  F_{[a,b]} \left( \bC \right) }}}\,.
	\label{eq:freeEnergyDef} 
\end{equation}
Note that we have simply employed eq. (\ref{klll}), valid for an arbitrary function $A(\bC)$ of the adjacency matrix. In order to calculate the average $\langle (\cdots) \rangle_{\rm ER}$ of the logarithm of the partition function, we use the replica method \cite{edwardsAnderson} 
\begin{equation}
	f = - \lim_{n \rightarrow 0} \lim_{N \rightarrow \infty} 
	\frac{1}{\beta N n}\ln\left(\frac{\avER{Z^n  e^{ yNF_{[a,b]} \left( \bC \right) }}} {\avER{e^{yNF_{[a,b]}  \left( \bC \right)  } }}\right)\,,
	\label{eq:logarithmReplica} 
\end{equation}
in which we have exchanged the order of the limits $n \rightarrow 0$ and $N \rightarrow \infty$. This assumption, although very difficult to prove in the general case, is usually harmless, and it allows us to compute the free energy in the  thermodynamic limit by solving a saddle-point integral. The general strategy of the replica approach consists firstly in evaluating the ensemble average in Eq. \eqref{eq:logarithmReplica} for a positive integer $n$. After the thermodynamic limit is taken, one considers $n \in \mathbb{R}$ and then continues $n$ analytically to $n \rightarrow 0$. Even though the replica method is generally a non-rigorous approach, it has a long tradition in the statistical physics of disordered systems as a correct heuristic method to evaluate ensemble averages \cite{BookParisi}.

Since all pairwise couplings in our model are ferromagnetic, exact results for the thermodynamics of the system are obtained by simply restricting ourselves to the replica symmetric solutions for the order parameter \cite{BookParisi,Monasson1998}. All the details of the replica calculation are explained in appendix \ref{Ising}. Here we just present the final analytical expression for the replica symmetric free energy per spin
\begin{widetext}
\begin{equation}
\begin{split}
\beta f=&
-c \mu_y^2 \ln 2 -\frac{c\mu_y^2}{2}
 \int d\theta d\theta' W(\theta) W(\theta')  
 \ln\left[\frac{\cosh(\beta J)}{
1+
\tanh (\beta \theta)
\tanh (\beta \theta')
\tanh (\beta J )
}\right]
\\
  &-\sum_{k=0}^\infty
  p_y(k)
\int\left[   \prod_{l=1}^k d\theta_l W(\theta_l)\right] 
\ln\left(
\frac{
	\cosh\left[ \beta\left(h + \beta^{-1} \sum_{l=1}^k  \arctanh[\tanh(\beta J) \tanh(\beta \theta_l)] \right)\right]
}
{2^{k-1}\prod_{l=1}^k \cosh\{\arctanh[\tanh(\beta J) \tanh(\beta \theta_l)]\}}
\right),
\end{split}
\label{eq:freeEnergyFinal} 
\end{equation}
\end{widetext}
where $W(\theta)$ is the distribution of effective local fields \cite{Monasson1998}, obtained from the solution of the self-consistent distributional equation
\begin{equation}
\begin{split}
	W(\theta)&=\sum_{k=0}^\infty  q_y(k)\int \left[ \prod_{l=1}^{k}  d\theta_l 
	W(\theta_l) \right]
		\\&\times
\delta\left(\theta-
h- \frac{1}{\beta} \sum\limits_{l=1}^{k} \arctanh[ \tanh(\beta J)\tanh(\beta \theta_l) ]
\right) .
\end{split}
\label{eq:WstarFinal} 
\end{equation}
The quantity $\mu_y$ in Eqs. \eqref{eq:freeEnergyFinal} and \eqref{eq:WstarFinal} encodes the microscopic graph structure of the constrained ensemble of ER graphs. The parameter $\mu_y$ is obtained from \cite{Isaac_Fernando}
\begin{equation}
\mu_y= \underset {\mu}{\operatorname {arg\max} }
	\{\mathcal{F}_y(\mu)\}
	=
\frac{
\sum\limits_{k=0}^\infty 
\mu_y^k p_c(k)
e^{y I_{[a,b]}( k+1)}
}
{
\sum\limits_{k=0}^\infty 
\mu_y^k p_c(k)e^{y I_{[a,b]}( k)}
}\,,
        \label{eq:muyDef}
\end{equation}
where the function $\mathcal{F}_y(\mu)$ reads
\begin{equation}
\mathcal{F}_y(\mu)
-\frac{c}{2}-\frac{ c\mu^2}{2}
+ \ln\left(\sum\limits_{k=0}^\infty e^{yI_{[a,b]}(k)}\mu^k p_c(k)\right)\,.
\label{jjjl}
\end{equation}
The quantities $p_y(k)$ and $q_y(k)$, appearing in Eqs. \eqref{eq:freeEnergyFinal} and \eqref{eq:WstarFinal}, are computed  from the following equations
\begin{equation}
	p_y(k)=\frac{\mu_y^k p_c(k)e^{y I_{[a,b]}( k)}
}
{\sum\limits_{q=0}^\infty 
\mu_y^qp_c(q)e^{y I_{[a,b]}( q)}}\,,
\label{eq:pcTilde} 
\end{equation}
\begin{equation}
	 q_y(k)=
\frac{
	\mu_y^k p_c(k)e^{y I_{[a,b]}( k+1)}
}
{\sum\limits_{q=0}^\infty 
\mu_y^qp_c(q)e^{y I_{[a,b]}( q+1)}}\,,
\label{eq:qcTilde} 
\end{equation}
with $k \in \{0,1,2,\dots\}$. The quantity $p_y(k)$ is the probability that a randomly chosen node has degree $k$, while $q_y(k)$ is the probability that a node at one of the extremes of a randomly chosen edge has degree $k+1$ \cite{Newman10}. Both quantities depend on $y$, since they refer to the constrained ensemble of graphs generated from Eq. \eqref{eq:probWeightedY}. By combining Eqs. \eqref{eq:muyDef} and \eqref{eq:pcTilde}, one obtains that $c \mu_y^2 = \langle k \rangle_y$, where
\begin{equation}
  \langle k \rangle_y = \sum_{k=0}^\infty k p_y(k)
  \label{meandeg}
\end{equation}
is the mean degree in the constrained ensemble.  We point out that Eq. \eqref{eq:WstarFinal} has no closed analytical solution in the general case and one has to resort to the population dynamics algorithm  \cite{MezardPopulationDynamics} in order to obtain a numerical solution to the distribution $W(\theta)$.

Our aim is to characterize the different phases of the Ising model and the nature of the transitions between them. Thus, it is interesting to calculate the intensive magnetization, obtained from the derivative of the free energy with respect to the external field $h$
\begin{equation}
	m=\int d\theta	 \tilde W(\theta) \tanh(\beta \theta)\,,
	\label{eq:mWtilde} 
\end{equation}
where the distribution $\tilde W(\theta)$ is determined from
\begin{equation}
\begin{split}
		\tilde W(\theta)&=
	\sum_{k=0}^\infty  p_y(k)
\int\left[   \prod\limits_{l=1}^k d\theta_l W(\theta_l)  \right] 
			     \\ &\times\delta\left(\theta-
				     h- \frac{1}{\beta} \sum\limits_{l=1}^{k} \arctanh[ \tanh(\beta J)\tanh(\beta \theta_l) ] \right)\,.
\end{split}
\label{eq:Wtilde}
\end{equation}
The derivative of $m$ with respect to $h$ yields the magnetic susceptibility
\begin{equation}
	\chi=\int d\theta	 \frac{\partial \tilde W(\theta)}{\partial h} \tanh(\beta \theta)\,,
	\label{eq:ChiWtilde}
\end{equation}
while the analytical expression for the intensive internal energy $u$ reads
\begin{equation}
\begin{split}
	u&=
-hm
-\frac{Jc \, \mu_{y}^2}{2}\bigg\{\tanh(\beta J)
       \\&{}-
 \int   
\frac{d\theta d\theta' W(\theta) W(\theta') \tanh (\beta \theta) \tanh (\beta \theta') \sech^2(\beta J)}{
 1+ \tanh (\beta \theta)
\tanh(\beta \theta')
\tanh(\beta J)  
}
\bigg\}\,.
\label{eq:EnergyW} 
\end{split}
\end{equation}

An important feature of the phase diagram of the Ising model is the critical inverse temperature $\beta_c$ where the system changes its behavior from ferromagnetic to paramagnetic in a continuous way. Since the moments of the order parameter distribution $W(\theta)$ vary continuously across this transition, we can use bifurcation analysis and derive the following equation for $\beta_c$ (see appendix \ref{Ising} for details)
\begin{equation}
\langle k\rangle_{q_y}\tanh(\beta_c J)=1\,,
\label{eq:criticalT} 
\end{equation}
with
\begin{align*}
\langle k\rangle_{q_y}=\sum_{k=0}^\infty q_y(k) k\,.
\end{align*}
Random graph models usually undergo a second-order percolation transition as a function of the average degree $c$ \cite{Newman10}. In particular, the largest connected component of ER random graphs contains a total number of $O(N)$ nodes provided $c > 1$. Since the $T \rightarrow 0$ limit of the magnetization of the Ising model on a random graph gives the fraction of nodes belonging to the giant connected component \cite{Leone2002}, the limit $T \rightarrow 0$ of Eq. \eqref{eq:criticalT} yields the critical line marking the {\it continuous} percolation transition in the constrained ensemble defined by Eq. \eqref{eq:probWeightedY}.

%%%%%%%%%%%%%%%%%%%%
\subsection{Numerical results}
\begin{figure}[!htbp]
	 \includegraphics[scale=0.75]{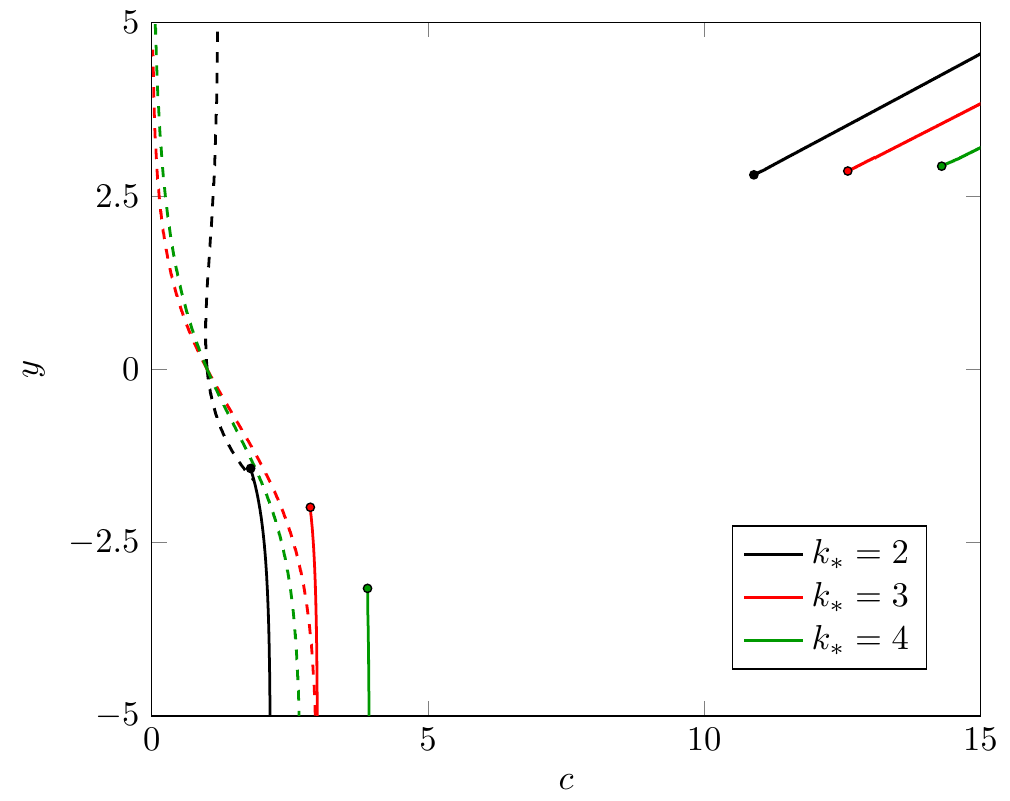}
         \caption{Phase diagram illustrating the second-order percolation transition (dashed lines) and the first-order condensation transition (solid lines) for the constrained ER random graph ensemble generated from eq. \eqref{eq:probWeightedY}, where the degrees are conditioned to lie in the interval $[k_{*}-1,k_{*}+1]$. For $k_{*} =2$, the second-order percolation transition terminates at a given value $y < 0$, below which it becomes discontinuous, coinciding with the condensation transition.}
         \label{fig:cartoon1}
 \end{figure}

%%%%%%%%%%%%%%%%%%%%
We start by discussing the phase diagram for the structural transitions in the constrained ensemble for different intervals $[a,b]$ controlled by a single parameter $k_{*}$, defined through $a = k_{*}-1$ and $b = k_{*}+1$. Figure \ref{fig:cartoon1} shows  the critical lines for the second-order percolation transition (dashed lines) and for the first-order condensation transition (solid lines) in the plane $(c,y)$ for different values of $k_{*}$. The continuous percolation transition is obtained by solving the equation $\langle k \rangle_y = 1$, derived from the limit $T \rightarrow 0$ of eq. (\ref{eq:criticalT}), while the condensation transition is obtained by finding the discontinuity of the fraction $f$ of nodes having degrees in $[a,b]$. 
  
 For fixed values of $c$, the critical values $y_{c}$ on the solid lines  identify the condensation transition: for $|y|> |y_{c}|$, the degree  distribution $p_y(k)$ is peaked on a few degrees, while $p_y(k)$ exhibits a Poisson-like behavior for $|y|<|y_{c}|$ \cite{Isaac_Fernando}.  For fixed values of $y$, the critical values $c_p$ on the dashed lines mark the percolation transition: for $c < c_p$, the graph is solely composed of finite connected components, whereas a giant connected component containing $O(N)$ nodes emerges for $c > c_p$.  As shown in figure \ref{fig:cartoon1}, for $k_{*} =2$, the continuous percolation transition meets the condensation transition at a certain value of $y$, below which the percolation transition becomes first-order. Thus, in the case of $k_{*} =2$, the solid line appearing for low $c$ identifies both the percolation and the condensation transition.

\begin{figure}[!htbp]
	 \includegraphics[scale=0.95]{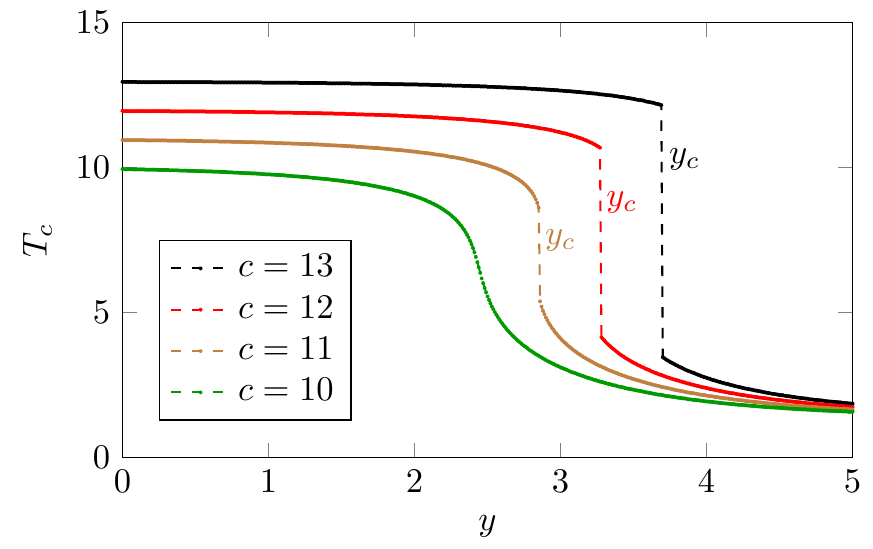}
         \caption{Critical temperature $T_c$ (see Eq. \eqref{eq:criticalT}) identifying the continuous phase transition between the paramagnetic and the ferromagnetic phases. The results for $T_c$ are shown as a function of $y$ for different values of the average degree $c$. The dashed lines mark the critical values $y_{c}$ at which the constrained ER ensemble undergoes a first-order structural transition to a phase exhibiting condensation of degrees (see Fig. \ref{fig:cartoon1}).}
         \label{fig:phaseDiagramIsing}
\end{figure}

Figure \ref{fig:phaseDiagramIsing} shows the critical temperature $T_c$ for the second-order phase transition between the paramagnetic and the ferromagnetic phases as a function of $y$. For sufficiently large $c$, $T_c$ drops discontinuously when $y$ crosses the condensation transition, which is a consequence of the abrupt decrease of the average degree $\langle k \rangle_y$ in the condensed phase.
 
Finally, we study the effect of condensation of degrees on the thermodynamics of the Ising model. As we approach the first-order condensation transition in the plane $(c,y)$ (see figure \ref{fig:cartoon1}), the function $\mathcal{F}_y(\mu)$ displays two maxima,  one of them being metastable.
 
\begin{figure}[!h]
  \includegraphics[scale=0.95]{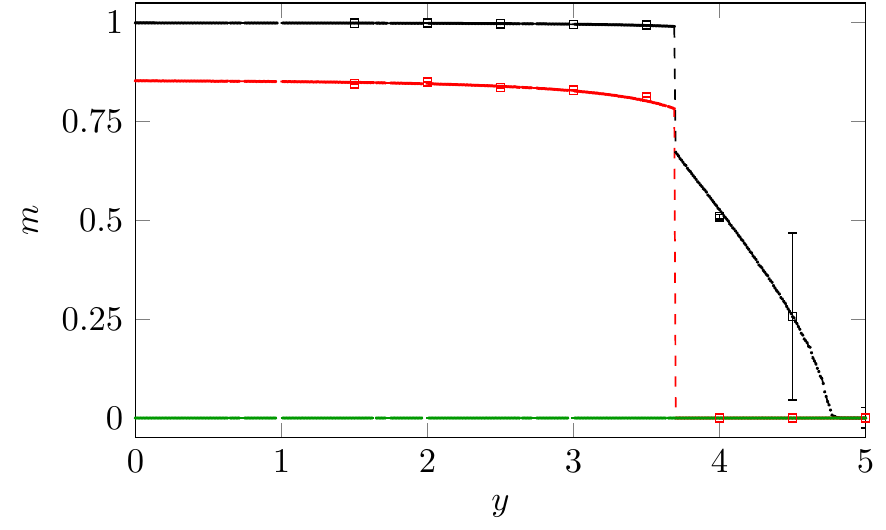} \\
  \includegraphics[scale=0.95]{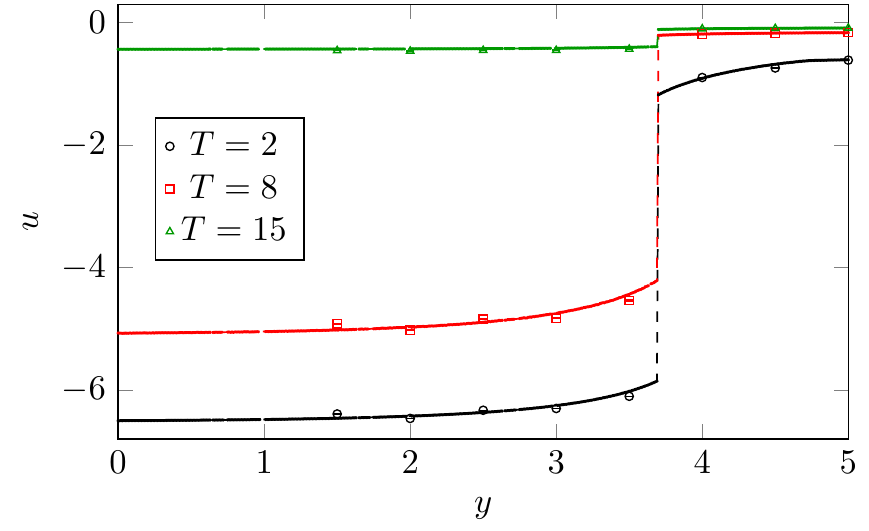} \\
   \includegraphics[scale=0.95]{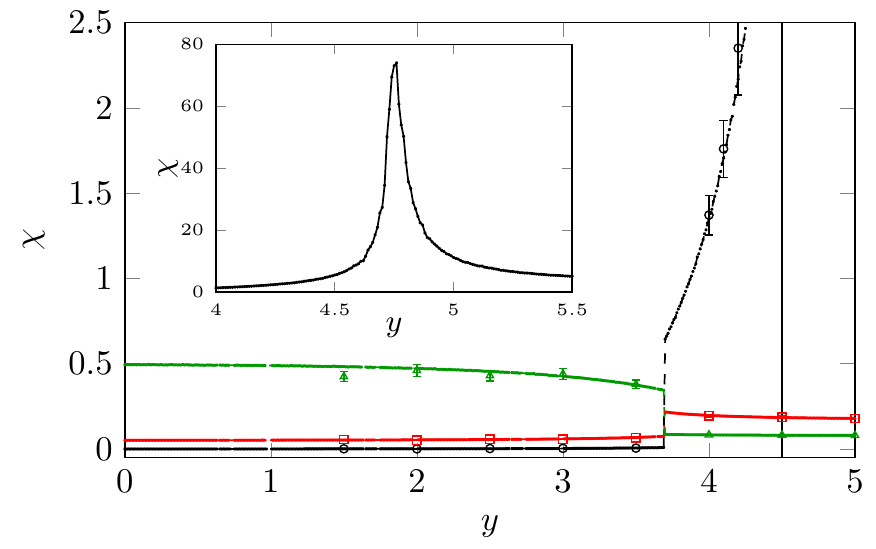}
\caption{Magnetization $m$, internal energy $u$, and magnetic susceptibility $\chi$ of the Ising model as a function of $y$ for average degree $c=13$, zero external magnetic field ($h=0$), and different temperatures $T$. The theoretical results (solid lines) are derived from the numerical solution of Eq. \eqref{eq:WstarFinal} using the population dynamics algorithm, while the different symbols are results obtained from Monte Carlo simulations of the model with a total number of $N=1000$ spins. The inset shows the behavior of $\chi$ around the second-order phase transition between the paramagnetic and the ferromagnetic phases for $T=2$.}
\label{magnetic} 
\end{figure}

 Figure \ref{magnetic} exhibits the magnetization, the internal energy, and the susceptibility for $c=13$. All quantities are shown as a function of $y$, for three different values of $T$, using figure \ref{fig:phaseDiagramIsing} as a guide. For $T=15$, even though the magnetization is always zero, the system exhibits a first-order transition between two paramagnetic phases at $y=y_c$, since the susceptibility and the internal energy display a jump at $y=y_c$. For $T=8$, the magnetization  drops to zero at $y_c$, while $u$ and $\chi$ increase discontinuously. Such behavior characterizes a first-order transition between a ferromagnetic and a paramagnetic phase. Finally, for $T=2$, the Ising model undergoes two different phase transitions as a function of $y$. Firstly, $m$, $u$ and $\chi$ varies discontinuously at $y=y_c$, with the magnetization changing between two finite values, which characterizes a first-order phase transition between ferromagnetic states. By further increasing $y$ in the regime $y > y_c$, we notice that $m$ vanishes continuously, while the magnetic susceptibility seems to diverge at a certain $y$. The latter behavior is typical of the usual second-order phase transition between ferromagnetic and paramagnetic states occurring in the Ising model.
 
%%%%%%%%%%%%%%%%%%%%
\section{Spectral properties of constrained random graphs}
\label{spectrumXX}
\subsection{The eigenvalue distribution}
%%%%%%%%%%%%%%%%%%%%
In this section we analyze the impact of the condensation transition on the eigenvalue distribution of ER random graphs drawn from Eq. \eqref{eq:probWeightedY}. By defining the eigenvalues $\lambda_1(\bC),\dots,\lambda_N(\bC)$ of a single instance of the symmetric adjacency matrix $\bC$, the empirical spectral distribution reads
\begin{equation}
\rho_N(\lambda)= \frac{1}{N}\sum_{i=1}^N \delta\left[ \lambda- \lambda_i(\bC) \right]\,.
\label{eq:spectralDensityDefinition} 
\end{equation}
Here we are interested in the average eigenvalue distribution corresponding to rare graph configurations labeled by $y$. Thus, following the prescription of eq. (\ref{klll}), we perform the ensemble average of $\rho_N(\lambda)$ over atypical regions of the ensemble space as follows
\begin{equation}
\rho_y(\lambda) = \lim_{N \rightarrow \infty} \frac{\avER{\rho_N(\lambda)  e^{y N  F_{[a,b]}\left( \bC \right) } } }{\avER{ e^{y N  F_{[a,b]}\left( \bC \right)}} }\,.
  \label{spec}
\end{equation}
The calculation of $\langle . \rangle_{\rm ER}$ in the above equation can be recasted in a problem  analogous to the computation of the average free energy in a spin-glass model \cite{Edwards}
\begin{equation}
	\rho_y(\lambda)=
        -\frac{2}{N\pi} \lim_{\eta\rightarrow0^+} {\rm Im} \left[ \,  \frac{\left\langle \partial_z \ln Z(z) e^{y N  F_{[a,b]}\left( \bC \right) } \right\rangle_{\rm ER}}
          {\left\langle e^{y N  F_{[a,b]}\left( \bC \right) } \right\rangle_{\rm ER}   } \right]\,,
\label{pol}
\end{equation}
where $Z(z)$ is the analogous of a partition function
\begin{align*}
Z(z)=\int_{-\infty}^{\infty} \left(  \prod_{i=1}^N dx_i \right) e^{-\frac{i}{2}\sum\limits_{i,j=1}^N x_i \left( z \delta_{ij} - c_{ij} \right) x_j}\,,
\end{align*}
with $z= \lambda-i\eta$ and $\partial_z \equiv \frac{\partial}{\partial z}$. The behavior of $\rho_y(\lambda)$ as a function of $y$ will allow us to characterize the effect of condensation of degrees on the global spectral properties of $\bC$.

The average spectral density $\rho_y(\lambda)$ can be computed using both the replica and the cavity methods, as developed in the context of sparse random matrix theory \cite{Edwards,Kuhn_2008,Isaac_2008}. Here we compute the ensemble average in Eq. \eqref{pol} by using the replica approach, whose main technical details are explained in appendix \ref{appspec}.  The analytical expression for $\rho_y(\lambda)$ is given by
\begin{align}
	\rho_y(\lambda)=-\lim_{\eta\rightarrow0^+} \frac{1}{\pi}
 \int d\Delta \, \tilde Q(\Delta) \, {\rm Im} \Delta ,
 \label{eq:rhoLambdaQTilde} 
\end{align}
where $\Delta \in \mathbb{C}$ and $d \Delta \equiv d {\rm Re} \Delta \, d {\rm Im} \Delta$. The joint distribution $\tilde Q(\Delta)$ of the real and imaginary parts of the complex variable $\Delta$ is determined from 
\begin{equation}
\begin{split}
	\tilde Q(\Delta)&=
	\sum_{k=0}^\infty p_y(k)
 \int  \left[\prod_{l=1}^k d\Delta_l Q(\Delta_l)\right]
		      \\& \times
 \delta\left(
\Delta+
	\frac{1}{z+\sum_{l=1}^k \Delta_l }
\right)\, ,
\end{split}
\label{kp}
\end{equation}
where $Q(\Delta)$ obeys the self-consistent equation
\begin{equation}
\begin{split}
	Q(\Delta)&=
	\sum_{k=0}^\infty q_y(k)
 \int  \left[\prod_{l=1}^k d\Delta_l Q(\Delta_l)\right]
		      \\& \times
 \delta\left(
\Delta+
	\frac{1}{z+\sum_{l=1}^k \Delta_l }
\right)\,.
\end{split}
\label{eq:QDeltaPopulationDynamics} 
\end{equation}
The quantities $p_y(k)$ and $q_y(k)$, determined respectively by Eqs. \eqref{eq:pcTilde} and \eqref{eq:qcTilde}, encode the statistical properties of the degrees in the constrained ensemble of graphs. The quantity $\tilde Q(\Delta)$ can be easily identified as the distribution of the diagonal elements of the resolvent matrix associated to $\bC$ \cite{Metz2010}.

In order to characterize the fluctuations of the eigenvalue distribution, it is interesting to consider the second moment of the spectral density
\begin{equation}
\langle \lambda^2 \rangle_{\rho_y}  =\int d\lambda \rho_y(\lambda)\lambda^2.
\end{equation}
One can easily show that
\begin{equation}
  \langle \lambda^2 \rangle_{\rho_y} = \langle k \rangle_y\,,
  \label{lp}
\end{equation}
where $\langle k \rangle_y$ follows from eq. (\ref{meandeg}). Since $\langle \lambda \rangle_{\rho_y} = 0$ due to the symmetry  $\rho_y(\lambda) = \rho_y(-\lambda)$, the variance of the distribution $\rho_y(\lambda)$ is fully determined by the average degree in the constrained ensemble.

%%%%%%%%%%%%%%%%%%%%
\subsection{Numerical results}
\begin{figure}[!h]
  \includegraphics[scale=0.80]{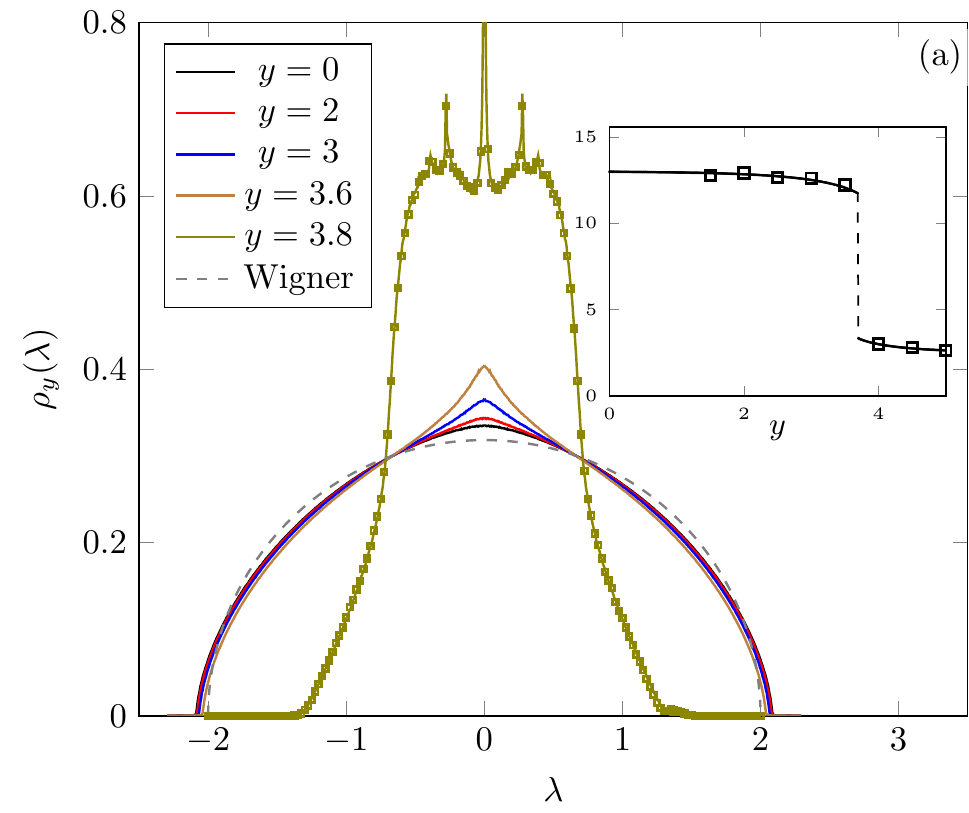} \\
  \includegraphics[scale=0.80]{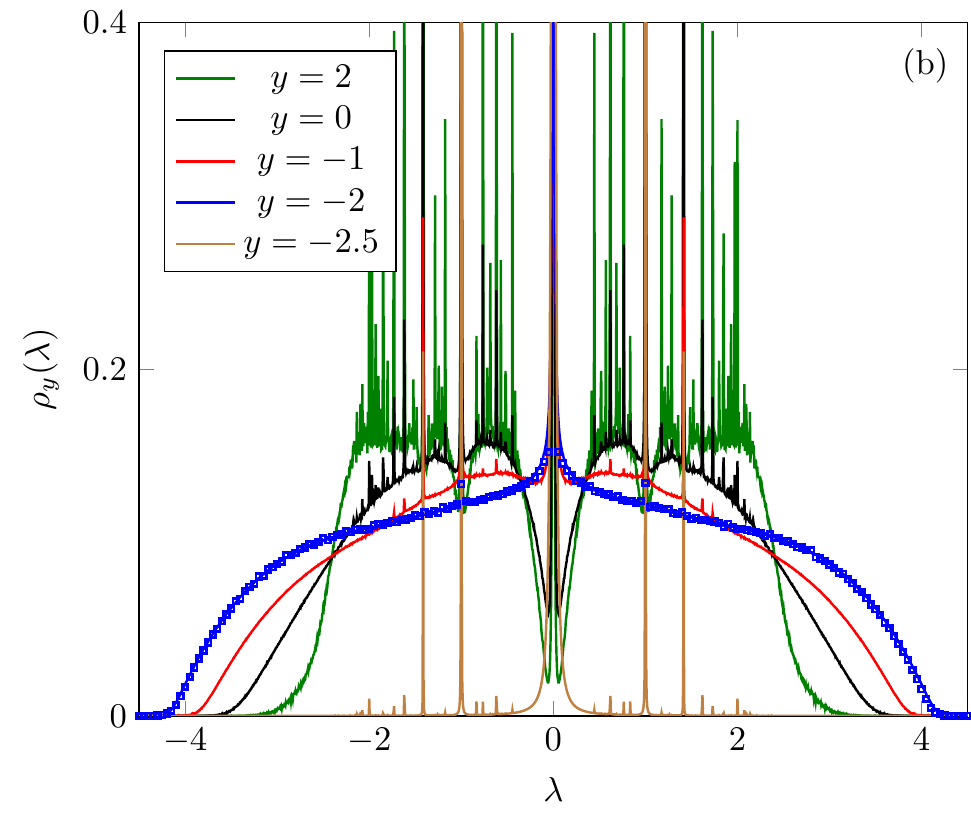}
  \caption{Theoretical results (solid lines) for the spectral density of constrained ER random graphs for different values of $y$ (see Eq. \eqref{eq:probWeightedY}), interval $[1,3]$, and average degrees (a) $c=13$ and (b) $c=2$. Figures (a) and (b) show the behavior of the eigenvalue distribution as we cross the condensation transition for high and low $c$, respectively.  The theoretical results are obtained from the numerical solution of Eq. \eqref{eq:QDeltaPopulationDynamics} using the population dynamics algorithm. The square symbols are obtained from the direct  diagonalization of $1000$ independent realizations of the $1000 \times 1000$ adjacency matrix characterizing atypical graph configurations generated through a reweighted Monte Carlo method \cite{Isaac_Fernando}. The eigenvalues have been rescaled as $\lambda_i \rightarrow \lambda_i / \sqrt{c}$ in subfigure (a). The inset shows the second moment of the spectral density $\rho_y(\lambda)$ for $c=13$.}
\label{fig:crossingPhaseTransitionSpectralDensity}
\end{figure}

%%%%%%%%%%%%%%%%%%%%
Here we discuss the outcome of solving Eqs. (\ref{eq:rhoLambdaQTilde}-\ref{eq:QDeltaPopulationDynamics}) numerically for different values of $y$ using the population dynamics method. The results are presented in figure \ref{fig:crossingPhaseTransitionSpectralDensity} for two fixed values of $c$, in order to capture the effect of the first-order condensation transition occurring at small and large average degrees (see figure (\ref{fig:cartoon1})).

For large $c$,  $\rho_y(\lambda)$ is approximately given by the Wigner semicircle law when $y=0$. By increasing $y$, $\rho_y(\lambda)$ gradually develops a bump at $\lambda=0$, until the eigenvalue distribution suddenly becomes more concentrated around $\lambda=0$ for $y > y_c$, which reflects the large fraction of degrees lying in the interval $[a,b]$ within the condensed phase. Accordingly, the variance of $\rho_y(\lambda)$ drops discontinuously as $y$ crosses the critical point $y=y_c$, as illustrated in the inset of figure \ref{fig:crossingPhaseTransitionSpectralDensity}(a).

For low values of $c$, the distribution $\rho_{y}(\lambda)$ corresponding to typical graph configurations ($y=0$) is composed of many delta peaks, most of them located at the eigenvalues of finite trees \cite{Bauer2001}. The delta peaks gradually disappear for decreasing $y <0$, until the distribution $\rho_y(\lambda)$ abruptly collapses into a few delta peaks when $|y| > |y_c|$. In particular, figure \ref{fig:crossingPhaseTransitionSpectralDensity}(b) suggests that, when $|y| > |y_c|$, the peak at $\lambda=0$ has the largest weight in comparison to the others. This feature is consistent with the degree distribution $p_y(k)$ characterizing the condensed phase appearing in this specific region of the phase diagram, where $p_y(k)$ displays a large peak at $k=0$ \cite{Isaac_Fernando}.

Overall, figure \ref{fig:crossingPhaseTransitionSpectralDensity} shows that the condensation transition leads to a dramatic change of the eigenvalue statistics. This is in contrast, for instance, to the standard second-order percolation transition, which does not bring about any qualitative changes in the moments of the spectral density \cite{Bauer2001}, even though the structure of the graph changes in a striking way.

%%%%%%%%%%%%%%%%%%%%
\section{Final remarks}
\label{conclusionXX}
%%%%%%%%%%%%%%%%%%%%
Random graphs undergo structural transitions when certain control parameters are changed. Here we have studied the effect of a discontinuous transition in the topology of Erd\H{o}s-R\'enyi (ER) random graphs on two different problems: the thermodynamic behavior of the Ising model defined on ER random graphs, and the eigenvalue statistics of the adjacency matrix of ER graphs. This structural transition identifies the discontinuous appearance of rare graph samples having a large number of nodes with similar degrees, following from an abrupt change in the degree statistics. We have shown that this condensation transition has a profound impact on the equilibrium properties of the Ising model as well as on the spectral properties of random graphs.

In the case of the Ising model, the condensation transition leads to a rich phase diagram, including additional first-order phase transitions between the paramagnetic and the ferromagnetic phases, which are absent in the typical equilibrium behaviour of the Ising model without an external magnetic field. We have characterized the transitions among the different phases in terms of the magnetization, the internal energy, and the magnetic susceptibility. Concerning the spectral properties of ER random graphs, we have shown that the condensation transition in the graph structure leads to a discontinuous behaviour of the eigenvalue statistics of the adjacency matrix. In particular, the variance of the eigenvalue distribution displays a jump at the condensation transition, which characterizes the abrupt change in the total number of edges. The exactness of our main theoretical results have been supported by Monte Carlo simulations. 

The first-order phase transitions discussed here are detected by varying a control parameter $y$, which is coupled to a random variable that counts how many degrees lie in an arbitrary interval $[a,b]$. Thus, $y$ enables to probe rare sectors of the graph ensemble space, since this parameter essentially controls the ``distance''  from the regime of typical fluctuations ($y=0$). Therefore, from the perspective of large deviation theory, the condensation transition in the degree statistics is triggered by large deviations in the graph structure. The parameter $y$ has a more concrete meaning when we interpret the generation of {\it rare} graph samples in the original model (see Eq. \eqref{klla}) as the generation of {\it typical} graph samples in a constrained ER ensemble (see Eq. \eqref{eq:probWeightedY}). In this setting, we can picture a variation in $y$ as a change in the graph topology, where some edges are rewired in order to comply with a certain average fraction of degrees in $[a,b]$. This is indeed one of the main ideas underlying the reweighted Monte Carlo approach to generate atypical ER graph samples \cite{Isaac_Fernando,Hart}.

Here we have illustrated the impact of condensation of degrees in a paradigmatic model of cooperative behavior, i.e. the Ising model, and on an important spectral observable for dynamical processes on graphs, i.e. the eigenvalue distribution of the adjacency matrix. Although the condensation transition is a statistically rare event, from the results reported here we expect that condensation of degrees has a striking effect on the macroscopic behavior of other large interacting systems modelled through random graphs. Thus, we hope our work stimulates the research towards a better understanding of the effects of condensation of degrees in different topics, such as synchronization phenomena on networks \cite{Doro2002}, diffusion processes on graphs \cite{Noh2004}, the linear stability of sparse interacting systems \cite{Neri2016,Neri2019}, and the dynamics of network formation \cite{Newman10}.

Finally, we point out that condensation of degrees is driven by weak correlations between the degrees of ER random graphs. Since the eigenvalues of a sparse random matrix are weakly correlated random variables  \cite{Metz_PerezC2017,PerezC_Metz2018,PerezC_Metz2018_2}, it would be  interesting to study whether these eigenvalues undergo a similar condensation transition.

\begin{acknowledgments}
I. P. C and F. L. M.  thank London Mathematical Laboratory for financial support. F. L. M. also acknowledges a fellowship and financial support from CNPq/Brazil (Edital Universal 406116/2016-4).
\end{acknowledgments}

\bibliography{references.bib}

\appendix
%%%%%%%%%%%%%%%%%%%%%%%%%%
\section{Analytical calculations for the Ising model}
\label{Ising}
%%%%%%%%%%%%%%%%%%%%%%%%%%
In this appendix we derive in detail the thermodynamical properties of  the Ising model on random graphs generated by Eq. \eqref{eq:probWeightedY}. 

%%%%%%%%%%%%%%%%%%%%%%%%%%
\subsection{The replica symmetric free energy}
%%%%%%%%%%%%%%%%%%%%%%%%%%
Firstly, we discuss how to compute the ensemble average $\avER{Z^n  e^{  yNF_{[a,b]} \left( \bC \right)  }}$ appearing in Eq. \eqref{eq:logarithmReplica}. By taking $n$  to be  a  positive integer we write
\begin{align}
\nonumber 
&\avER{Z^ne^{yN F_{[a,b]} \left( \bC \right)  }}
=\sum_{k_1,\dots,k_N = 0}^{N-1}  \sum_{\bm{\sigma}_1,\dots,\bm{\sigma}_n} \exp \left(\sum\limits_{i=1}^N \mathcal H_i \right)\\
&\times\avER{\exp\left(\beta J \sum\limits_{i<j} c_{ij}\sum\limits_{a=1}^n\sigma_{ia} \sigma_{ja}\right)\prod_{i=1}^N \delta_{K_i,k_i}}\,
\end{align}
where $\bm{\sigma}_a$ for $a=1,\ldots,n$ is the Ising vector for the $a$-th replica,  $K_i = \sum_{j =1 (\neq i)}^N c_{ij} $, and 
\begin{align}
\mathcal H_i= h\beta\sum\limits_{a=1}^n \sigma_{ia} +y I_{[a,b]} (k_i).
\end{align}
By rewriting the Kronecker delta functions in the Fourier representation, we obtain, after some algebra
\begin{equation}
\begin{split}
&\avER{\exp\left(\beta J \sum\limits_{i<j} c_{ij}\sum\limits_{a=1}^n\sigma_{ia} \sigma_{ja}\right)\prod_{i=1}^N \delta_{K_i,k_i}} \\
=&\int\left( \prod_{i=1}^N \frac{du_i}{2\pi}\right) \exp\Bigg\{i\sum\limits_{i=1}^N  u_ik_i \\
&{}+\frac{c}{2N}\sum\limits_{i,j} \left(\exp\left[\beta J \sum\limits_{a=1}^n\sigma_{ia} \sigma_{ja} -i  (u_i+u_j) \right]-1\right)\Bigg\}\,,
\end{split}
\label{eq:zneyAverageReplica} 
\end{equation}
where we  have already dropped subextensive terms, which are unimportant in the thermodynamic limit. Next, we define the spin vectors in the replica space $\underline{\sigma}_i=(\sigma_{i1},\dots,\sigma_{in})$ for  $i=1,\dots,N$, and we introduce the following order parameter function 
\begin{equation}
P(\underline\sigma)=\frac{1}{N} \sum_{i=1}^N e^{-i u_i}\delta_{\underline \sigma,\underline \sigma_i}\,.
\label{eq:parameterOrderIsing} 
\end{equation}
After some algebra we are left with the following expression 
\begin{equation}
\avER{Z^ne^{yR_{[a,b]}}}= \int D\{P,\hat P\}\, e^{N S(P,\hat P)}\,,
\end{equation}
where $\int D\{P,\hat P\}$ denotes a path integral over the pair  $\{P,\hat{P}\}$, and
\begin{equation}
\begin{split}
S(P,\hat P) &=\ln\left(\sum\limits_{\underline{\sigma}}  e^{ h\beta \sum\limits_{a=1}^n \sigma_a} \sum\limits_{k=0}^\infty  \frac{(-i)^k  }{k!}(\hat  P(\underline\sigma))^k e^{y I_{[a,b]}( k)} \right)\\&{}-\frac{c}{2}+\frac{c}{2}\sum\limits_{\underline\sigma, \underline \tau}P(\underline\tau)P(\underline\sigma)e^{\beta J \underline \sigma \cdot \underline\tau }
+i  \sum\limits_{\underline \sigma} \hat  P(\underline\sigma)P(\underline\sigma)\,.
\end{split}
\label{eq:ZnIntegralPP} 
\end{equation}
In the thermodynamic limit, this path integral can be evaluated by using the saddle-point method, at which the pair  of functions $\{P,\hat{P}\}$ obeys the following saddle-point equations
\begin{align}
&-i   \hat  P(\underline\sigma)= c \sum\limits_{\underline \tau} P(\underline\tau)e^{\beta J \underline \sigma \cdot \underline\tau }\label{eq:saddleEqF1}\,, \\
&P(\underline \tau) =\frac{ e^{ h\beta \sum_{a=1}^n \tau_a} \sum_{k=1}^\infty  \frac{(-i\hat  P(\underline\tau))^{k-1}}{(k-1)!} e^{y I_{[a,b]}( k)}}{\sum\limits_{\underline{\sigma}} e^{ h\beta \sum_{a=1}^n \sigma_a}\sum_{k=0}^\infty \frac{(-i \hat P(\underline\sigma))^k}{k!}e^{y I_{[a,b]}( k)}}\,.
\label{eq:saddleEqF2} 
\end{align}

%%%%%%%%%%%%%%%%%%%%%
\subsection{Replica symmetric ansatz}
\label{sub:replica_symmetric_ansatz}
%%%%%%%%%%%%%%%%%%%%%
Within  replica symmetric ansatz we assume the functions $P$ and $\hat P$ to take the following form:
\begin{equation}
\begin{split}
P(\underline \sigma)&=\mu_y \int d\theta ~ W(\theta) \prod_{a=1}^n \frac{e^{\beta \theta \sigma_a}}{2 \cosh(\beta \theta)}\,,\\
-i \hat P(\underline \sigma)&= c \mu_y \int du~ H(u) \prod_{a=1}^n \frac{e^{\beta u \sigma_a}}{2 \cosh(\beta u)}\,,
\label{eq:pphatRS} 
\end{split}
\end{equation}
where  $W(\theta)$ and $H(u)$ are  densities yet to be determined. Notice that  the constant  $\mu_y$ in Eq. \eqref{eq:pphatRS} also needs to be determined, but fairly naturally, this will turn out to be precisely the  factor $\mu_y$  given by  Eq. \eqref{eq:muyDef}.\\
 By using the replica symmetric ansatz in Eqs. \eqref{eq:saddleEqF1} and \eqref{eq:saddleEqF2} we obtain, after some algebra 
\small
\begin{equation}
H(u)=\int d\theta W(\theta) \delta\left(u-\frac{1}{\beta}\arctanh[ \tanh(\beta J)\tanh(\beta \theta)]\right),
\label{eq:QuSaddle} 
\end{equation}
and,
\begin{align}
W(\theta)&=\sum\limits_{k=0}^\infty q_y(k) \int \left[\prod_{l=1}^{k}  du_l H(u_l) \right] \delta\left(\theta-h-\sum\limits_{l=1}^{k} u_l\right)\,,
\label{eq:muWtheta0} 
\end{align}
where $q_y(k)$ follows Eq. \eqref{eq:qcTilde}.  This gives back Eq. \eqref{eq:WstarFinal} in the main text.\\
Similarly, we can evaluate the expression of $S(P,\hat P)$ within the replica symmetric ansatz yielding
\begin{equation}
\begin{split}
&\tilde{\mathcal F}(W,H)=-\mathcal F(y,\mu_y) -\frac{nc\mu_y^2}{2} \ln[\cosh(\beta J)]\\
&+n\left\{c \mu_y^2   \int d\theta du W(\theta) H(u)   \ln \left( \frac{\cosh(\beta (u +\theta))}{2 \cosh(\beta u)\cosh(\beta \theta)} \right)
\vphantom{\sum\limits_{k=0}^\infty} \right. \\
&{}-\frac{c\mu_y^2}{2}\int d\theta d\theta' W(\theta) W(\theta')  \ln(1+\tanh\beta \theta \tanh\beta \theta' \tanh\beta J)\\
 &- \left.\sum\limits_{k=0}^\infty \! p_y(k) \!\!
\int\!\left[   \prod\limits_{l=1}^k du_l H(u_l)\right] 
\!\ln\!\left( \frac{ \cosh\left[ \beta\left(h + \sum_{l=1}^k u_l \right)\right] } {2^{k-1}\prod_{l=1}^k \cosh(\beta u_l )} \right)\!\! \right\}\\
&+\mathcal{O}(n^2)\,.
\end{split}
\label{eq:freeEnergyFinal2}
\end{equation}
\normalsize
As shown in \cite{Isaac_Fernando}, $p_y(k)$ is the effective probability distribution for the degree of a node, which can be used to rewrite \eqref{eq:muyDef} as follows,
\begin{align}
\nonumber 
\mu_y&=\frac{1}{c\mu_y}\sum_{k=0}^\infty(k+1) p_y(k+1) = \frac{1}{c\mu_y} \avW{k}\,,
\end{align}
that is, $c \mu_y^2$ is the average of the degree of a node. The derivation of expressions for magnetization, internal energy, and bifurcation analysis to obtain the critical temperature follow the standard route.

%%%%%%%%%%%%%%%%%%%%%%%%%%
\section{Derivations for the spectral density}
\label{appspec}
%%%%%%%%%%%%%%%%%%%%%%%%%%
We start by noticing that the expression for the spectral density given by Eq. \eqref{eq:spectralDensityDefinition} is mathematically  similar to the one  in Eq. \eqref{eq:freeEnergyDef}, corresponding to the Ising model. By using the replica method we write
\begin{equation}
\begin{split}
\rho_y(\lambda)&=-\lim_{n\rightarrow0} \lim_{\eta\rightarrow0^+} \frac{2}{Nn\pi}{\rm Im} \,\partial_z \ln\left(\frac{\avER{Z^n(z)  e^{yNF_{[a,b]}}}} {\avER{e^{yNF_{[a,b]}} }}\right) \,,
\end{split}
\label{eq:rhoLambdaReplicaSymmetric} 
\end{equation}
where the numerator can be worked out to obtain
\begin{equation}
\begin{split}
&\avER{Z^n(z)  e^{yN F_{[a,b]}}}= \sum_{k_1,\dots,k_N}\int\left[\prod_{a=1}^n d\bm x_a^N\right] \exp\left(\sum_{j=1}^N \mathcal H_j\right)\\
&\times\int \left[\prod_{i=1}^N \frac{du_i}{2\pi}\right]\exp\Bigg\{i \sum\limits_{i=1}^N u_i k_i\\ 
& \hspace{0.5 cm}{}+\frac{c}{2N}\sum\limits_{i,j} \left(\exp\left[i \sum\limits_{a=1}^n x_{ia} x_{ja}-i  (u_i+u_j)\right]-1\right)\Bigg\}\,,
\end{split}
\end{equation}
with the definition
\begin{align}
\mathcal H_j=-\frac{iz}{2} \sum\limits_{a=1}^n  x^2_{ja}+y  I_{[a,b]}(k_j)\,.
\end{align}
Next, we introduce the following functional order parameter
\begin{align*}
P(\underline x)=\frac{1}{N}\sum_{i=1}^N \delta(\underline x-\underline x_{i})e^{-i u_i}\,,
\end{align*}
with $\underline x_i=(x_{i1},\dots,x_{in})$. This allows us to write the following path integral
\begin{equation}
\avER{Z^n(z)  e^{yR_{[a,b]}}} = \int D\{P,\hat P\} e^{N S(P,\hat P)},
\end{equation}
with
\begin{equation}
\begin{split}
S(P,\hat P)&= \ln \int d\underline x \,e^{ -\frac{iz}{2} \sum\limits_{a=1}^n x_a^2} \sum\limits_{k=0}^\infty \frac{(-i\hat P(\underline x))^k}{k!}  e^{ y  I_{[a,b]}(k) }\\
&-\frac{c}{2}+\frac{c}{2}\!\int\! d\underline x d\underline y\,P(\underline x)P(\underline y)e^{i\underline x\cdot \underline y}\\
&+i \int d \underline x\,  P(\underline x)\hat P(\underline x)\,.
\end{split}
\label{eq:ZnP} 
\end{equation}
The asymptotic behavior of the path integral is evaluated by  the saddle point method, which yields a set of couple saddle-point equations for $P$ and $\hat{P}$.\\ 
In this case the replica symmetric ansatz can be written as follows
\begin{align}
\nonumber 
P(\underline x)&=\mu_y\int d\Delta  Q(\Delta )\prod_{a=1}^n \sqrt{\frac{1}{2\pi i \Delta}} e^{i \frac{x^2_a}{2\Delta}}\,,\\
-i \hat P(\underline x)&=c \mu_y\int  d\Gamma \Omega(\Gamma)\prod_{a=1}^n \sqrt{\frac{\Gamma}{2\pi i}} e^{i \Gamma \frac{x^2_a}{2}}\,,
\end{align}
where the densities $Q(\Delta )$ and $\Omega(\Gamma)$ are determined by plugging this ansatz into the saddle-point equations. The latter become
\begin{equation}
\begin{split}
\Omega(\Gamma)&= \int d\Delta W(\Delta) \delta\left(\Gamma+\Delta\right)\,,\\
Q(\Delta)&=\sum\limits_{k=0}^\infty q_y(k) \int\! \left[ \prod_{l=1}^k d\Gamma_l \Omega(\Gamma_l)\right]  \delta\!\left(\!\Delta+\frac{1}{z-\sum_{l=1}^k \Gamma_l }\right)\,,
\end{split}
\end{equation}
which, when combined,  yield Eq. \eqref{eq:rhoLambdaQTilde} reported in the main text.\\
Finally, one can show that using the replica symmetric ansatz in Eq.  \eqref{eq:ZnP}, the spectral density $\rho_y(\lambda)$ given by Eq. \eqref{eq:rhoLambdaReplicaSymmetric}, becomes
\begin{equation}
\begin{split}
\rho_y(\lambda)&= \lim_{\eta\rightarrow 0+} \frac{1}{\pi} {\rm Im} \sum\limits_{k=0}^\infty i p_y(k) \\
& \int  \left[\prod_{l=1}^k d\Gamma_l \Omega(\Gamma_l)\right] \frac{ \int dx \, x^2 e^{ -\frac{i}{2}  x^2\left(z-\sum_{l=1}^k\Gamma_l \right)}}{ \ \int dx \,  e^{ -\frac{i}{2}  x^2\left(z-\sum_{l=1}^k\Gamma_l \right)}}\\
&= \lim_{\eta\rightarrow 0+} \frac{1}{\pi} {\rm Im} \int  d\Delta  \frac{ \int dx \, ix^2 e^{ i\frac{x^2}{2\Delta}  }}{ \ \int dx \,  e^{i \frac{ x^2}{2\Delta} }}\\\
& \sum\limits_{k=0}^\infty p_y(k) \int  \left[\prod_{l=1}^k d\Gamma_l \Omega(\Gamma_l)\right]\delta\left(\Delta +\frac{1}{z-\sum_{l=1}^k\Gamma_l }\right)\,,
\label{eq:rhoLambdaBeforeGaussians} 
\end{split}
\end{equation}
where $p_y(k)$ is defined in Eq. \eqref{eq:pcTilde}. 

\end{document}